\documentclass{elsart3}
\usepackage[square,comma]{natbib}
\usepackage{graphicx}
\usepackage{amssymb}
\usepackage{epsfig}
\usepackage{verbatim}

\newcommand{\bea}{\begin{eqnarray}}
\newcommand{\be}{\begin{equation}}
\newcommand{\ee}{\end{equation}}
\newcommand{\eea}{\end{eqnarray}}
\newcommand{\bfi}{\begin{figure}}
\newcommand{\efi}{\end{figure}}
\newcommand{\bc}{\begin{center}}
\newcommand{\ec}{\end{center}}
\newcommand{\nn}{\nonumber}

\newtheorem{teo}{Theorem}

\begin{document}
\begin{frontmatter}

\title{Exact Results for the Roughness of a Finite Size Random Walk}
\author[fis,fis3]{V. Alfi},
\corauth[cor]{Corresponding author}
\ead{valentina.alfi@roma1.infn.it}
\author[fermi]{F. Coccetti},
\author[afs]{M. Marotta},
\author[cnr]{A. Petri},
\author[fis,afs]{L. Pietronero}
\address[fis]{``La Sapienza'' University, Physics Department, \\
  P.le A. Moro 5, 00185, Rome, Italy}
\address[fis3]{``Roma Tre'' University, Physics Department, \\
  V. della Vasca Navale 84, 00146, Rome, Italy}
\address[cnr]{Istituto dei Sistemi Complessi - CNR, Via Fosso del Cavaliere 100, 00133 Roma, Italy}
\address[fermi]{Museo Storico della Fisica e Centro Studi e Ricerche ``Enrico Fermi``, Via Panisperna, Roma, Italy}
\address[afs]{Applied Financial Science, New York, USA}

\begin{abstract}
We consider the role of finite size effects on the value of the effective
Hurst exponent $H$. This problem is motivated by the properties of the high
frequency daily stock-prices. For a finite size random walk
we derive some exact results based on Spitzer's identity.
The conclusion is that finite size effects strongly enhance the value of
$H$ and the convergency to the asymptotic value ($H=1/2$) is 
rather slow. This result has a series of conceptual and practical
implication which we discuss.
\end{abstract}


\end{frontmatter}

\section{Introduction}

The dynamics  of stock-prices can be described as a 
subtle form of random walk with complex properties and correlations. 
The main characteristics which are usually considered are the broad
distribution of price returns (fat tails) and the clustering of 
volatility \citep{bib1,bib2,bib3}.
Recently, also the roughening properties, defined by the Hurst 
exponent $H$ \cite{bib4}, have been considered \cite{bib5,bib6,bib7}.
The idea is that the roughening exponent can provide information which 
goes beyond the above two properties. However, the extraction of 
this information is not simple and the usual interpretation of deviation
from the random walk  value ($H=1/2$) is in terms of generic long range
correlations.
Sometimes the Hurst exponent is simply used as a generic statistical
indicator without particular interpretation \cite{bib8}.
Here we show that finite size effects in the real data can alter significantly
the determination of the effective Hurst exponent.
In fact, high frequency stock-price data are relatively homogeneous
within the same day, but the large night jump (often of the order of the total
daily fluctuation), 
implies an intrinsic limit on the size of the dataset \cite{bib14}. Motivated by these
observations we consider how finite size effects affect the determination
of the Hurst exponent for a finite size random walk. We derive exact results
for the expectation  value of the maximum value of a random walk using the
Spitzer's identity \cite{bib12}.
From this one can derive the effective Hurst exponent for the case in which
the vertical fluctuation is defined by the maximum and the minimum values.
There are also different methods to estimate $H$ but we believe that,
especially for financial data, the one based on the maximum and the minimum
values is especially relevant because of the role played by these values in
various trading strategies. \\
The main result is that the effective Hurst exponent is strongly
enhanced by finite size effects and that the convergency to the
asymptotic value $H=1/2$ is rather slow. This result will
be further enhanced by the inclusion of ``fat tails'' and non-stationary
properties.
This implies that high frequency daily stock-prices are unavoidably
affected by finite size effects \cite{bib14}. In addiction non-stationarity can make 
the convergency much slower providing a possible alternative interpretation
of the deviations from $H=1/2$ which are observed in long time series.

\section{Exact analysis of a Finite Size Random Walk}
In this section we derive some exact result for finite Random Walks which
are necessary in order to consider the roughness properties for a finite
size system.
Suppose that ${\delta x_1, \delta x_2, ..., \delta x_n}$ are 
independent random variables, each taking the value $+1$ with probability $p$,
and $-1$ otherwise.
Consider the sums:
\bea 
X_n=\sum_{i=1}^{n}\delta x_i
\eea
then the sequence $X=\{X_i:i \geq 0 \}$ is a simple random walk starting at the
origin.
In order to compute the expectation value of the maximum and the minimum
of the walk after $n$ steps, is useful to consider the following 
theorem.

\begin{teo}[Spitzer's identity]{\upshape{\cite{bib12}}}
Assume that $X$ is a right-continuous random walk, and let 
$M_n=\max\,\{X_i: 0\geq i \geq n \}$ be the maximum of the walk
up to time $n$. Then, introducing the auxiliary variables $s$ and $t$, for
$|s|, |t| <1$ one has {\upshape{Eq.\ref{spitzer}}},
\bea
\log \Big(\sum_{n=0}^{\infty}\,t^n \,\mathbb E(s^{M_n})\Big)=
\sum_{n=1}^{\infty}\frac 1n\, t^n \, \mathbb E \,(s^{X_n^+})        
\label{spitzer}
\eea
where $X_n^{+}=\max\, \{0,X_n\}$ and $\mathbb E$ is the expectation value.
\end{teo}
Considering the exponential of both  member of Eq.\ref{spitzer} one has:

\bea
\sum_{n=0}^{\infty}\,t^n \,\mathbb E(s^{M_n})=
\exp\Big(\sum_{n=1}^{\infty}\frac 1n\, t^n \, \mathbb E \,(s^{X_n^+})\Big)
\label{spitzer2}
\eea
The $k$-derivative with respect to $t$ of the left hand side of
Eq.\ref{spitzer2} for $t=0$, gives:
\bea
\frac{{\partial}^k}{\partial t^k}{\Bigg \vert}_{0}=k!\;\mathbb{E}(s^{M_k})
\label{lefts}
\eea
Defining the right side member of Eq.\ref{spitzer2} as $f(t)$  one can derive:
\bea
\nn
\frac{{\partial}^k}{\partial t^k}f(t){\Bigg \vert}_{0}&=&
\sum_{j=1}{k}
\frac{(k-1)!}{(k-j)!}\mathbb{E}(s^{X_j^+})\Bigg(\frac{{\partial}^{(k-j)}}{\partial  t^{(k-j)}}f(t)\Bigg){\Bigg \vert}_{0}\\
f(0)&=&1
\label{rights}
\eea
By equating Eq.\ref{lefts} and Eq.\ref{rights} one obtains:
\bea 
\mathbb{E}(s^{M_k})=\frac{1}{k}\sum_{j=1}^k\frac{\mathbb{E}(s^{X_j^+})}{(k-j)!}\Bigg(\frac{{\partial}^{(k-j)}}{\partial  t^{(k-j)}}f(t)\Bigg){\Bigg \vert}_{0}
\label{final}
\eea
In order to obtain $\mathbb{E}(M_k)$ from the function $\mathbb{E} (s^{M_k})$
it is useful to consider the following expansion which holds for a symmetrical
probability density function:
\bea
&{\mathbb {E}}& (s^{X_j^+})=\int_{-\infty}^{\infty}s^{X_j^+}P_j(X_j)\,dX_j\\
\nn
&=&\int_{-\infty}^{0}P_j(X_j)\,dX_j+\int_{0}^{\infty}s^{X_j}P_j(X_j)\,dX_j\\
\nn
&=&\frac 12+\int_{0}^{\infty}s^{X_j}P_j(X_j)\,dX_j\\
\nn
&\simeq&\frac 12+\int_{0}^{\infty}(1+X_j\ln(s)+\frac 12 X_j^2 {(\ln(s))}^2+...)P_j(X_j)\,dX_j\\
\nn
&\simeq&1+\ln (s)\int_{0}^{\infty}X_jP_j(X_j)\,dX_j + {{\mathcal O}} {(\ln(s))}^2\\
\nn
&=&1+\frac 12 \,\mathbb E (|X_j|)\,\ln(s)+{\mathcal O} {(\ln(s))}^2
\eea
We now insert this result into Eq. \ref{final}. Considering also the 
identity:
\bea
\mathbb E (M_k)=\lim_{s\rightarrow 1}\frac{\mathbb E (s^{M_k})-1}{\ln(s)}
\eea
we finally obtain:
\bea
\mathbb E (M_k)=\sum_{i=1}^{k}\frac{\mathbb E (|X_i|)}{2i}
\label{calcolo}
\eea
Now we consider various possibilities for the specific nature of the 
random walk:
\begin{description}
\item[a)]  If the increments $\delta x$ are  independent and corresponding to 
a gaussian distribution with $\mathbb{E}(\delta x)=0$ and variance $\sigma^2
=1$, one obtains:
\bea
\mathbb E (|X_i|)&=&\sqrt{\frac{2i}{\pi}}
\eea
\item[b)]If $\delta x$ have values $\pm 1$ with equal probability one gets:
\bea
\mathbb E (|X_i|)&=&\sum_{X_i=-i}^i |X_i|P_i(X_i)
\eea
where
\bea
P_i(X_i)={i  \choose {\frac{X_i+i}{2}}}{\Bigg (\frac 12\Bigg )}^i
\eea
This leads to 
\bea
\mathbb E (|X_i|)&=&\Bigg\{
\nn
\begin{array}{ll}
\frac{(i-1)!!}{(i-2)!!}\qquad\mbox{if $i$ is even}\\
\frac{(i)!!}{(i-1)!!}\qquad \mbox{if $i$ is odd}
\end{array}
\eea
\end{description}
These explicit results permit now to compute the exact expectation for finite
size random walk properties.
Note that Eq.\ref{calcolo} has a general value with the only assumption that
the increments are not correlated and symmetrically
distributed.
This means, for example, that one could test the properties of stock-prices for
finite size samples being able to separate the role of fat tail, included in
Eq.\ref{calcolo}, from the role of correlations.
In the present paper we use Eq.\ref{calcolo} to estimate the effective roughness exponent of finite size systems.

\section{Effective roughness for a finite size Random Walk}
\begin{figure}
  \begin{center}
    \begin{tabular}{cc}
      \resizebox{80mm}{!}{\includegraphics{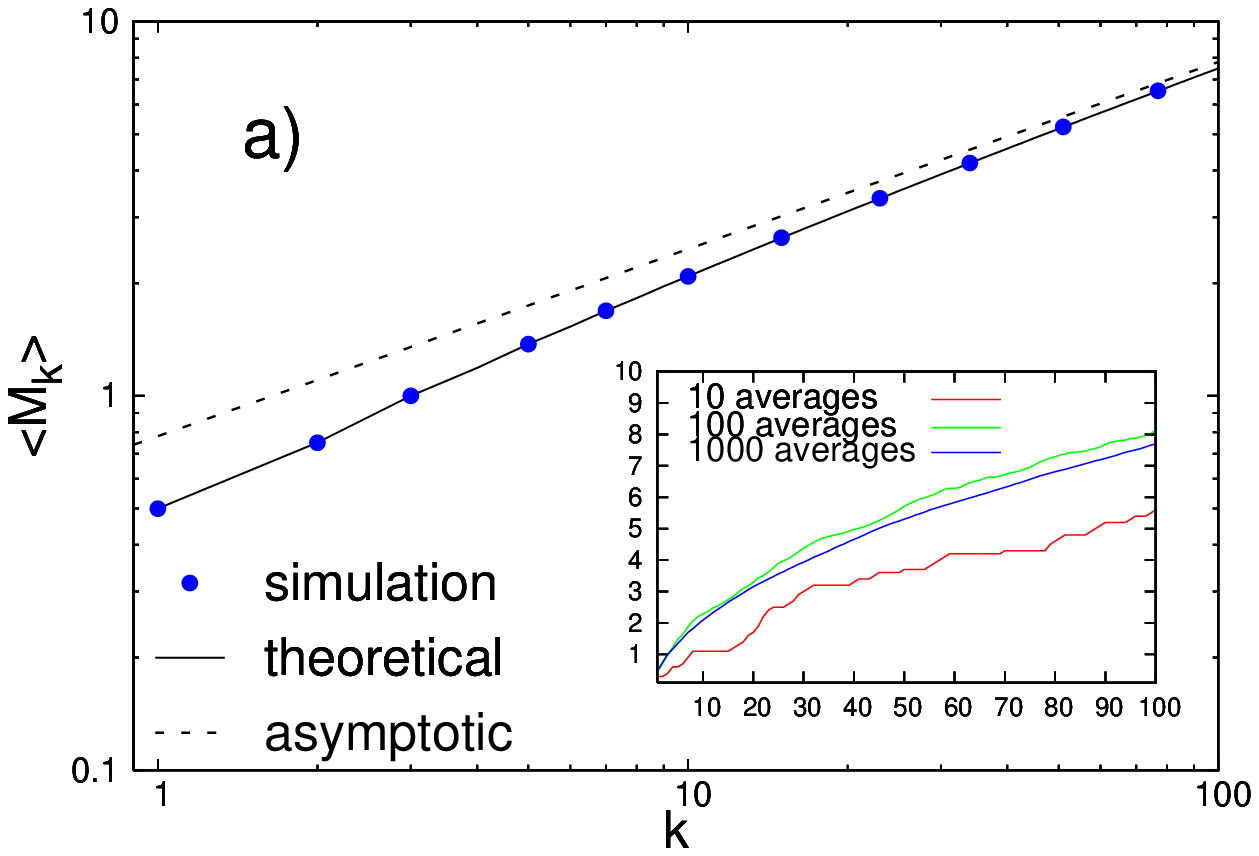}} & \\
      \resizebox{80mm}{!}{\includegraphics{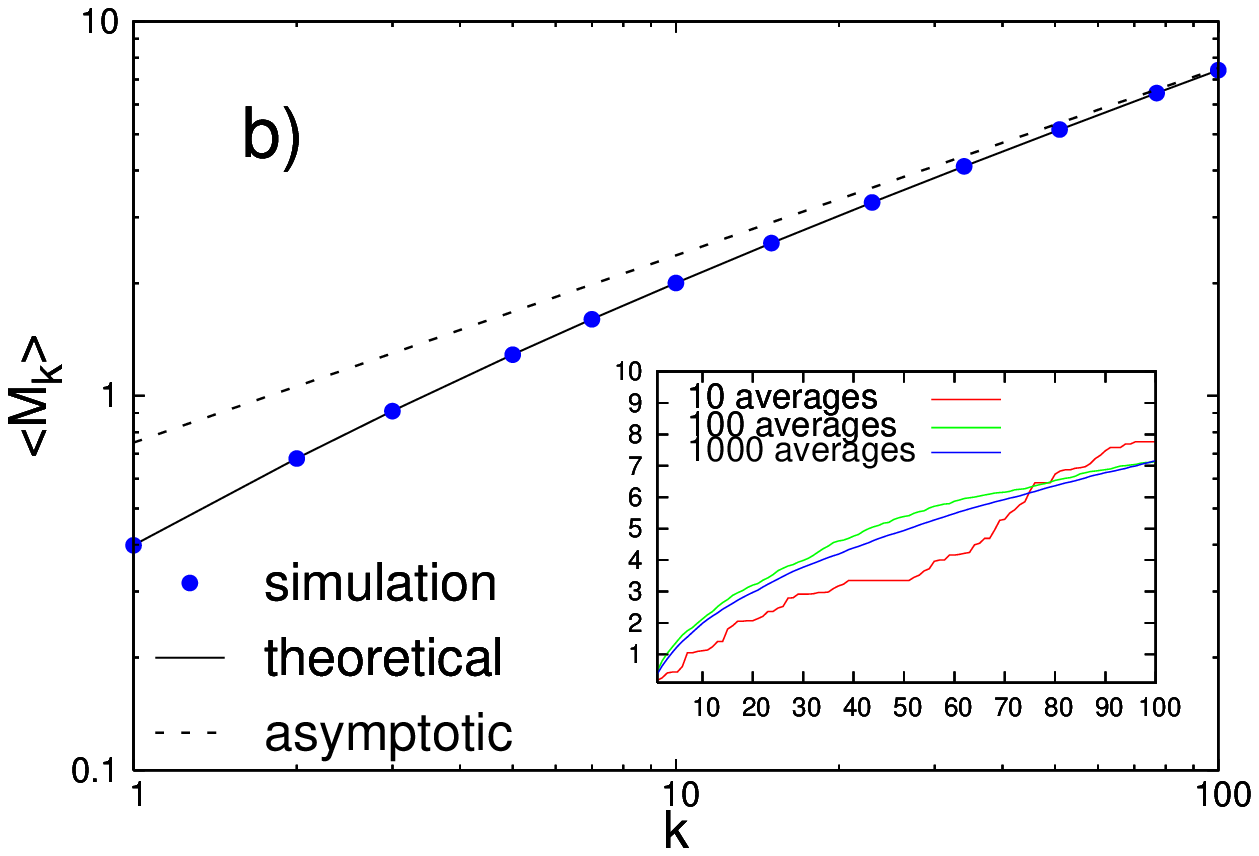}}&
    \end{tabular}
    \caption{The log-log graph shows $\mathbb E (M_k)$ as a function of $k$
      in the case of a) random walk with two identical steps  and b)  Gaussian random walk.
      In the inserts we show the convergency of the simulations to 
      expected value as a function of the number of realizations considered.}
    \label{Ngraf3}
  \end{center}
\end{figure}

Equation Eq.\ref{calcolo} gives an exact result relation between
the expectation value of the maximum value $M_k$ of a symmetric random walk of
$k$ steps, for a given probability distribution $\mathbb {E}(|X_i|)$
of the individual step. In terms of Monte Carlo simulations this
would correspond to an infinite number of samples. Since the Monte Carlo method
will be applied also to cases for which the analytical result is not 
available, we can use the present case as a test for the convergency of the
Monte Carlo method.\\
This comparison is shown in Fig.\ref{Ngraf3} (a and b) where the two 
inserts show precisely the degree of convergency as a function of 
the samples considered.
\\
In order to estimate the effective roughness exponent as a function of
the size of the interval considered, we can compute the maximum fluctuation
in a given interval of size $n$:
\bea
R(n)=\mathbb{E}(\max(n))-\mathbb{E}(\min(n))
\eea
and estimate the effective Hurst exponent $H$  from the scaling relation:
\bea
R(n)\simeq n^{H(n)}
\eea
For $n\rightarrow\infty$ we expect to recover the standard random walk
result $H(n\rightarrow\infty)=1/2$.

\begin{figure}[h]
  \begin{center}
    \resizebox{80mm}{!}{\includegraphics{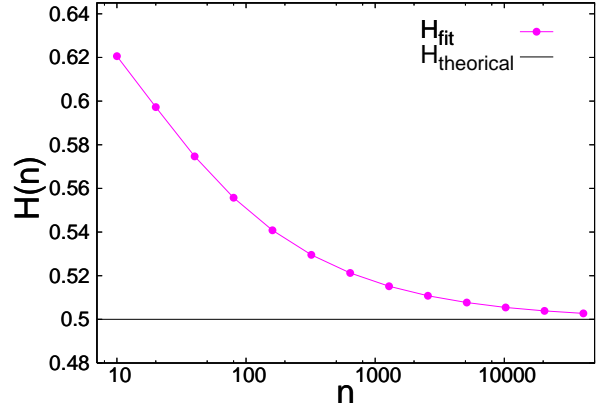}} 
    \caption{This plot shows the trend obtained fitting the curve 
${\mathcal  R}(n)$ for different values of the size $n$. The 
results shows an overestimate of Hurst exponent for small size, due to finite
size effects. This is a general result and it shows that finite size effects
always enhance the apparent Hurst exponent. This enhancement can be understood
by considering that, in some sense,a single step would correspond to $H=1$, so
the asymptotic value $H=1/2$ is approached from above.}
    \label{Ngraf5}
  \end{center}
\end{figure}

In Fig.\ref{Ngraf5} we report the effective Hurst exponent $H(n)$  as a function of
the  size $n$  of the interval considered. One can see that for finite 
value of $n$ the value of $H(n)$ is always larger than the asymptotic
value $H=1/2$ and that the convergency to the asymptotic value is rather slow.

This result has a number of implications:
\begin{description}
\item[(i)]
Usually a deviation of the value of $H$ from $1/2$ is interpreted in
terms of long range correlations \cite{bib8}.
We can see that a positive deviation $H>1/2$ can instead be due to finite size
effects. This result is especially relevant for high frequency stock-prices
data. In this case in fact the data are statistically homogeneous only within
a single day because the night jump is usually very large \cite{bib14}.
The typical number of transactions, for stocks of intermediate volatility,
ranges between $500$ and $3000$. In order to have
a statistical significance the maximum interval considered for the estimate of
$H$ should not exceed one tenth of their total number. The interval if scaled
to be considered ranges therefore  from a few transactions to about $100$ \cite{bib14}.
From Fig.\ref{Ngraf5} we can see that this would correspond to appreciable
deviation from $H=1/2$.\\
\item[(ii) Fat Tails and Short Range Correlations.]
Usually stock-price dynamics does not show short range
correlations. However, in case these would be present, their effect 
would be to decrease the effective number of independent steps, enhancing 
therefore the of finite size effects. A much more important implication
of the same type is due to the fat tails distributions of returns. 
In fact, in presence of a broad distribution of step size, the few
large steps will play a major role and the finite size effects will be
strongly enhanced. We are going to see that this point is very 
relevant for the analysis of high frequency stock-price data \cite{bib14}.
\item[(iii) Non-stationarity.]
The present analysis of finite size effects on the roughness is performed
under the hypothesis of a stationary process. It is well known instead that 
economic data show marked deviation from stationarity. This implies that the 
convergency to the asymptotic value $H=1/2$ can be much slower if one includes
these effects. In this perspective even data which refer to very long 
series may not reach convergency due to non stationarity. This implies 
a possible alternative origin for the deviation from $1/2$ which have been
reported for long time series\cite{bib14}.  
In the future we intend to consider specific models to test this possibility.
\end{description}


\begin{thebibliography}{99}

\bibitem{bib1} 
B.~Mandelbrot, Fractals and Scaling in Finance, Springer Verlag, New York, 1997.
\bibitem{bib2} 
R.N.~Mantegna, H.E.~Stanley, An Introduction to Econophysics, Cambridge University Press, Cambridge, 2000. 

\bibitem{bib3}
J.P.~Bouchaud, Theory of Financial Risk, Cambridge University Press, Cambridge, 2000.

\bibitem{bib4}
H.E.~Hurst, Long-term  storage capacity reservoirs, {\em Transaction of the   American Society of Civil Engineers 116, 770-808, 1951.}

\bibitem{bib5}
S.O.~Cajueiro, B.~Tabak, The Hurst exponent over time: testing the assertion that emerging market are becoming more efficient, {\em Physica A}, vol. 336, pag. 521-537, 2004.

\bibitem{bib6}
D.~Grech, Z.~Mazur, Can one make any crash prediction in finance using the local Hurst exponent idea?, {\em Physica A}, vol. 336, pag. 133-145, 2004.

\bibitem{bib7}
A.~Carbone, G.~Castelli, H.E.~Stanley, Time-dependent Hurst exponent in financial time series, {\em Physica A}, vol. 344, pag. 267-271, 2004.



\bibitem{bib8}
T.~Di Matteo, T.~Aste, M.M.~Dacorogna, Long-term memories of developed and emerging markets: Using the scaling analysis to characterize their stage of development, {\em Journal of Banking and Finance}, vol. 29, pag. 827-851, 2005.


\bibitem{bib12}
G.~Grimmet, D.~Stirzaker, Probability and Random Processes, Oxforf University Press, Oxford, 2001.



\bibitem{bib14} 
V.~Alfi, F.~Coccetti, M.~Marotta, A.Petri, L.Pietronero, Roughness and Finite Size Effect in the NYSE Stock-Price Fluctuations, printing in 2006.




\end{thebibliography}
\end{document}